\documentclass[11pt]{amsart}
\usepackage{amsmath}
\usepackage{amssymb}

\begin{document}
\newtheorem{Th}{Theorem}[section]
\newtheorem{Def}{Definition}[section]
\newtheorem{Lemm}{Lemma}[section]
\newtheorem{Cor}{Corollary}[section]
\newtheorem{Rem}{Remark}[section]
\newtheorem{Ex}{Example}[section]
\newtheorem{Conj}{Conjecture}
\newtheorem{Prob}{Problem}
\newtheorem{Propos}{Proposition}

%

\title{Functional Equations and the Generalised Elliptic Genus}

\author{H.~W. Braden~$^\dag$ and K.~E.~Feldman~$^\ddag$}

\address{$^\dag$ School of Mathematics, University of Edinburgh,\\
Mayfield Road, Edinburgh, Scotland, EH9 3JZ\\
~~E-mail: hwb@ed.ac.uk\\
$^\ddag$ DPMMS, University of Cambridge,\\ Wilberforce Road,
Cambridge, England, CB3 0WB\\
~~E-mail: k.feldman@dpmms.cam.ac.uk}

\date{}

\begin{abstract}We give a new derivation and characterisation of the generalised elliptic genus
of Krichever-H\"ohn by means of a functional equation.
\end{abstract}

\maketitle
\section*{Introduction}
Functional equations provide a common thread to several
investigations in mathematics and physics: our focus in this
article will particularly be on the areas of topology and
integrable systems where it is still unclear whether the threads
before us form part of a greater fabric. In topology the German
and Russian schools applied functional equations powerfully to
formal group laws and genera~\cite{BBNY,Bu1,BKh1,BKh2,BV,F,Hi,H,Hohn}. They have arisen
in the study of integrable systems in several different ways.
F.~Calogero, whom we honour in this volume, instigated in
\cite{Ca2} a new use of functional equations in the study of
integrable systems that is relevant here.

The modern approach to integrable systems is to utilise a Lax
pair. Calogero in \cite{Ca2}, by assuming a particular ansatz for
a Lax pair, reduced the consistency of the Lax pair to a
functional equation and algebraic constraints. In this way he
discovered the elliptic Calogero-Moser model. Similarly, by
assuming an ansatz for a realisation of the generators of the
Poincar\'e algebra, Ruijsenaars and Schneider \cite{RS} reduced
the ensuing commutation relations to that of a functional
equation. The Ruijsenaars-Schneider model which results from one
solution to this functional equation is also integrable. For the
Ruijsenaars-Schneider systems Bruschi and Calogero constructed a
Lax pair, again by means of an ansatz and consequent functional
equation \cite{BCa, BCb}. (The general solutions to the functional
equations of Ruijsenaars and Schneider have now been constructed
\cite{BBS,BSB}, but it is still open whether the resulting models
are completely integrable.) Later Braden and Buchstaber
generalised these various Lax pair ans\"atze \cite{BB1} and
encountered a rather ubiquitous functional equation \cite{BB} that
includes many functional equations arising in both cohomological
computations and integrable systems. We will return to this
functional equation in due course but what is of interest at this
stage is that the same equations arise in both settings. This may
reflect something deeper. String theory physics allows some
topology changes (such as flops) \cite{AGM, Witten2} and physical
quantities such as the partition function should reflect this
invariance; invariance under classical flops characterises the
elliptic genus \cite{To}. The authors of \cite{KSU} draw
connections between the complex cobordism ring and conformal field
theory. Certainly the Baker-Akhiezer functions associated to the
integrable systems satisfy addition formulae \cite{BK1, BK2} and
reflect the underlying algebraic geometry \cite{DFS}.

The present article aims to provide a new derivation of the
generalised elliptic genus of Krichever-H\"ohn by means of a
functional equation encountered in the study of integrable
systems. In the first section we will review equivariant genera of
loop spaces. The following section derives the relevant functional
equation which we then solve in the final section. Various remarks
will be made enroute that relate this approach to existing
derivations.

\section{Equivariant Genera of the Loop Space}

Motivated by the problem of obtaining left--right asymmetric
fermions in a Kaluza--Klein theory Witten in~\cite{Witten1}
suggested the study of a special twisted Dirac operator on closed
spin manifolds equipped with a smooth $S^1$-action. Witten
conjectured that the character--valued index of such a twisted
operator is in fact a constant and that the genus of a manifold
corresponding to this Dirac operator possesses  a rigidity
property.

 To break the conjecture into simpler pieces, Landweber posed a problem on
computation of a special ideal in the bordism ring of semifree $S^1$-actions
on spin manifolds. As a tool for the solution to this problem
Ochanine~\cite{Och} introduced an elliptic genus
$$
Q(x)=\frac{1}{2}\frac{x}{\mathrm
{tanh}(x/2)}\cdot\prod^{\infty}_{n=1} \left(\frac{(1+q^n\mathrm
e^x)(1+q^n\mathrm e^{-x})} {(1-q^n\mathrm e^x)(1-q^n\mathrm
e^{-x})}\cdot \frac{(1-q^n)^2}{(1+q^n)^2}\right).
$$

Meanwhile Witten~\cite{Witten} gave an informal approach to his own
conjecture by computing the equivariant signature of the space of smooth loops
${\mathcal L}M$ on the manifold $M$ and discovered that the genus obtained is
(up to a constant) equal to the Ochanine genus.

Although the equivariant signature of the loop space is not a well
defined object in algebraic topology, the formal properties of the
genus coming out of this procedure have many nice features. In
particular, the general methods of the theory of elliptic
operators and fixed point theorems~\cite{AB,AH} allowed Witten to confirm
his earlier conjecture in the reformulation that the elliptic genus, being formally equal to the
index of a Dirac type operator on the loop space, is rigid. This
statement was rigorously proved by Taubes~\cite{BT}.

Following Witten's general scheme one can calculate other well
known genera of the loop space. Consider the Hirzebruch
$\chi_y$-genus given by power series:
$$
R(x)=\frac{x\left(1+y\mathrm e^{-x(1+y)}\right)}{1-\mathrm e^{-x(1+y)}}.
$$

Let $M^{2n}$ be a stable almost complex manifold. Consider the canonical
$S^1$-action on the loop space ${\mathcal L}M^{2n}$ induced from the standard
$S^1$-action on the parameters:
$$
g:S^1\times {\mathcal L}M^{2n}\to {\mathcal L}M^{2n},\qquad
g(z,\gamma(t))=\gamma(zt), \qquad z,t\in S^{1},\qquad \gamma:S^1\to M^{2n}.
$$
The fixed point set of this action consists of constant loops only and,
therefore, is equal to $M^{2n}$. The explicit form of the
restriction of the tangent bundle $T({\mathcal L}M^{2n})$ to the loop
space ${\mathcal L}M^{2n}$ on $M^{2n}\subset {\mathcal
L}M^{2n}$ at a point $p\in M^{2n}$ is given by
$$
T_p({\mathcal L}M^{2n})\cong \Gamma(S^1\times T_pM^{2n})={\mathcal L}(T_pM^{2n}),
$$
where $\Gamma(S^1\times T_pM^{2n})$ is a space of sections of the bundle
$S^1\times T_pM^{2n}\to S^1$. From here the decomposition of
$T({\mathcal L}M^{2n})|_{M^{2n}}$ into eigenspaces with respect to the
$S^1$-action $g:S^1\times {\mathcal L}M^{2n}\to {\mathcal L}M^{2n}$ is
\begin{equation}
\label{dec}
T({\mathcal
L}M^{2n})\big|_{M^{2n}}=\sum^{\infty}_{k=-\infty}q^kTM^{2n},
\end{equation}
where $q=e^{2\pi it}$ acts on the $k$-th Fourier coefficient of
the loop $\gamma: S^1\to T_p(M^{2n})$ as multiplication by
$q^k$~(see details, for example, in~\cite{Hi}). The Atiyah--Bott
fixed point theorem says that a genus $\phi$ of an almost complex
manifold $X$ with a compatible circle action is equal to the sum
over all connected components $X_s$ of the fixed point set of
expressions
$$
\frac{\phi\left(T(X)|_{X_s}\right)}{e(\nu_s)},
$$
where $e(\nu_s)$ is the Euler class of the normal bundle $\nu_s$
of the embedding $X_s\subset X$, $s=1,2,...$. Applying it to the
decomposition into eigenspaces~(\ref{dec}) and $\phi=\chi_y$
we obtain formally
\begin{equation}
\chi_y({\mathcal L}M^{2n}) =\langle
\prod^m_{i=1}\left(\frac{x_i\left(1+y\mathrm e^{-x_i(1+y)}\right)}
{1-\mathrm e^{-x_i(1+y)}} \prod^{\infty}_{k=1}\frac{1+y\tilde q^k\mathrm
e^{-x_i(1+y)}}{1-\tilde q^k\mathrm e^{-x_i(1+y)}} \cdot\frac{1+y\tilde q^{-k}
\mathrm
e^{-x_i(1+y)}}{1-\tilde q^{-k}\mathrm e^{-x_i(1+y)}} \right), [M^{2n}]\rangle,
\label{ychi}
\end{equation}
where $\tilde q$ is now a formal parameter corresponding to the
generator of $H^*(CP(\infty),\mathbb Q)$. As the expression on the
right hand side of (\ref{ychi}) is not convergent, we rearrange
terms in such a way that we may rewrite
$$
\frac{1+y\tilde q^{-k}
\mathrm
e^{-x_i(1+y)}}{1-\tilde q^{-k}\mathrm e^{-x_i(1+y)}}=
\frac{\tilde q^k
\mathrm
e^{x_i(1+y)}+y}{\tilde q^k\mathrm e^{x_i(1+y)}-1}.
$$
Normalising the last expression by $-y^{-1}$ we arrive as in~\cite{Hi,Hohn} to
\begin{Def}
\label{CHI}
For a stable almost complex manifold $M^{2n}$ the equivariant $\chi_y$-genus
of the loop space ${\mathcal L}M^{2n}$ is defined up to a normalisation by
\begin{equation}
\chi_y({\mathcal L}M^{2n}) =\langle
\prod^m_{i=1}\left(\frac{x_i\left(1+y\mathrm e^{-x_i(1+y)}\right)}
{1-\mathrm e^{-x_i(1+y)}} \prod^{\infty}_{k=1}\frac{1+y\tilde q^k\mathrm
e^{-x_i(1+y)}}{1-\tilde q^k\mathrm e^{-x_i(1+y)}} \cdot\frac{1+y^{-1}
\tilde q^k
\mathrm
e^{x_i(1+y)}}{1-\tilde q^k\mathrm e^{x_i(1+y)}} \right), [M^{2n}]\rangle,
\label{chiy}
\end{equation}
for $\tilde q=\mathrm e^{2\pi i\tau}$, ${\mathrm {Im}}\tau>0$.
\end{Def}

\

Definition~\ref{CHI} describes a genus of stable almost complex
manifolds which is, up to a constant, equal  to the Krichever
genus~\cite{Kri} that was discovered as a particularly elegant
example of the theory of Conner--Floyd equations in complex
cobordism for circle actions~\cite{Kri1,Kri2,Kri3}. The theory of
Conner--Floyd equations allows one to deduce powerful restrictions
on the cobordism classes of stable complex manifolds with
compatible circle actions by means of the theory of complex
analytic functions. In particular, the rigidity property of a
genus is equivalent to a convergence of special analytic
expressions associated with the fixed point data of the circle
action on the manifold. The Krichever genus also has a rigidity
property but for a class of manifolds equipped with an
$SU$-structure.

Following general techniques one can relate the question of the
rigidity of any genus to the question of a multiplicative property
for some fibre bundles. The multiplicative property of a genus $g$
for the fibre bundle $p:E\stackrel{F}\longrightarrow B$ with
smooth fibre and base says that
$$
g(E)=g(F)\cdot g(B).
$$
It was shown by H\"ohn that the only genus $\phi$ that satisfies
the multiplicative property with respect to the fibre bundles
whose fibres admit an $SU$-structure is the Krichever genus. A
particular example of such bundles plays a crucial role in the
work of Totaro~\cite{To} who proved that the Krichever genus is
the only genus preserved by flops. The idea of Totaro was that the
difference (in the complex bordism ring of a point) of two complex
manifolds equivalent via a flop is bordant to a manifold $E$,
which is fibred over a certain complex manifold, and whose fibre
is $CP(3)$, equipped with a fake stable almost complex structure
that admits an $SU$-structure. Using the fact that the complex
bordism ring $\Omega^U_*$ has no torsion and that the flop is a
symmetric operation it was concluded in~\cite{To} that
$\phi$-genus of $CP(3)$ with such a stable almost complex
structure is zero. Thus, from the multiplicative property
\begin{equation}
\phi(E)=\phi(\overline {CP}(3))\,\phi(B)
\label{MULTCP}
\end{equation}
it follows that two manifolds equivalent via a flop have the same
$\phi$ genus. The proof in~\cite{To}, that there is no other genus
preserved by flops (or equivalently, 
satisfying $\phi(E)=0$ for any fibre bundle 
$E\stackrel{\overline {CP}(3)}\longrightarrow B$),
consists of an estimate on the size of the quotient of
$\Omega^U_*\times\mathbb Q$ by bordism classes of the differences
between manifolds equivalent via flops.

In the next section we rewrite the multiplicative property~(\ref{MULTCP}) in
terms of a functional equation for the generating function of
$\phi$. In the final section we will solve this functional
equation, giving a new derivation that $\phi$ is the
Krichever-H\"ohn genus.

\section{The Functional equation}

Let $M^{2n}$ be a stable almost complex compact manifold without
boundary. Consider complex vector bundles $\xi$ and $\eta$ over
$M^{2n}$ of complex dimension 2. Let $CP(\xi\oplus\eta)$ be the
complex projectivization of the Whitney sum $\xi\oplus\eta$, that
is an associated fibre bundle over $M^{2n}$ with fibre $CP(3)$. We
introduce a stable almost complex structure on $CP(\xi\oplus\eta)$
in the following way. Observe that
$$
T(CP(\xi\oplus\eta))\stackrel{\mathbb R}\cong \tau_F(CP(\xi\oplus\eta))\oplus
p^*T(M^{2n}),
$$
where $p:CP(\xi\oplus\eta)\to M^{2n}$ is the projection,
$\tau_F=\tau_F(CP(\xi\oplus\eta))$ is the bundle of tangents along
the fibre, and as usual, $T(X)$ denotes the real tangent bundle to
a manifold $X$. It is well known that for the complex
projectivization of any complex vector bundle $\eta$
\begin{equation}
\label{1spl}
\tau_F(CP(\eta))\stackrel{\mathbb R} \cong
{\mathrm {Hom}}_{\mathbb C}(\eta(1),\eta^{\bot})\cong
\eta^*(1)\otimes \eta^{\bot},
\end{equation}
where $\eta(1)$ is the tautological vector bundle over $CP(\eta)$,
$\eta\sp*(1)$ is conjugate to $\eta(1)$, and $\eta^{\bot}$ is its
orthogonal complement in $p^*{\eta}$:
$$
\eta(1)\oplus\eta^{\bot}\stackrel{\mathbb C}\cong p^*\eta.
$$
Adding the trivial complex line bundle $[1]_{\mathbb C}$
to the left hand side of~(\ref{1spl}) we obtain
\begin{align*}
\tau_F(CP(\xi\oplus\eta))\oplus[1]_{\mathbb C}& \stackrel{\mathbb
R} \cong  {\mathrm {Hom}}_{\mathbb C}(\eta(1),\eta^{\bot})\oplus
{\mathrm {Hom}}_{\mathbb C}(\eta(1),\eta(1))\\ &\stackrel{\mathbb
C} \cong {\mathrm {Hom}}_{\mathbb C}(\eta(1),p^*(\xi\oplus\eta))\\
&\stackrel{\mathbb C}\cong  \eta^*(1)\otimes p^*\xi \oplus
\eta^*(1)\otimes p^*\eta.
\end{align*}
Let us equip the
bundle $\tau_F(CP(\xi\oplus \eta))$ with a stable almost complex
structure in the following way
\begin{equation}
\tau_F(CP(\xi\oplus\eta))\oplus[1]_{\mathbb C}\stackrel{\mathbb
C}\cong
\eta^*(1)\otimes p^*\xi \oplus \left(\eta^*(1)\otimes p^*\eta\right)^*
\stackrel{\mathbb
C}\cong
\eta^*(1)\otimes p^*\xi \oplus \eta(1)\otimes p^*\eta^*.
\label{tauf}
\end{equation}
We define the stable almost complex structure on the
projectivization of $\xi\oplus\eta$ as the sum of the complex
structure (\ref{tauf}) and the standard complex structure in
$p^*T(M^{2n})$. To underline that this complex structure is not
the usual one, we denote the total space of the projectivization
with such a stable almost complex structure by $\overline
{CP}(\xi\oplus\eta)$.

\begin{Th}
A complex cobordism  genus $\phi:\Omega^U_*\to \mathbb Q$
satisfies the multiplicative property
$$
\phi(\overline {CP}(\xi\oplus\eta))=\phi(\overline
{CP}(3))\,\phi(M^{2n})
$$
if and only if its generating power series
$f(x)=1+\alpha_1x+\alpha_2 x^2+\dots$ is a solution of the
following functional equation:
\begin{equation}\label{funl}
\begin{split}
\lambda&=\frac{f(x_2-x_1)}{x_2-x_1}\cdot\frac{f(x_1-y_1)}{x_1-y_1}\cdot
\frac{f(x_1-y_2)}{x_1-y_2}+
\frac{f(x_1-x_2)}{x_1-x_2}\cdot\frac{f(x_2-y_1)}{x_2-y_1}\cdot\frac{f(x_2-y_2)}{x_2-y_2}\\
&\qquad -
\frac{f(x_1-y_1)}{x_1-y_1}\cdot\frac{f(x_2-y_1)}{x_2-y_1}\cdot\frac{f(y_1-y_2)}{y_1-y_2}-
\frac{f(x_1-y_2)}{x_1-y_2}\cdot\frac{f(x_2-y_2)}{x_2-y_2}\cdot\frac{f(y_2-y_1)}{y_2-y_1}
\end{split}
\end{equation}
for some constant $\lambda$.
\end{Th}
\begin{proof} By the same arguments as in~\cite{Hi} it is
sufficient to show that under the Gysin map $p_!:H^*(\overline
{CP}(\xi\oplus\eta),\mathbb Q)\to H^{*-6}(M^{2n},\mathbb Q)$ the
cohomology class $f(\tau_F)=f(\gamma_1)\cdots f(\gamma_4)$ (where
$\gamma_i$, $i=1,\dots4$, are the Chern roots of $\tau_F$) is
mapped into $H^0(M^{2n},\mathbb Q)$. To check it in our case we
can consider $BT^4=CP(\infty)\times CP(\infty)\times
CP(\infty)\times CP(\infty)$ as a base space instead of the
manifold $M^{2n}$, and we can put
$$
\xi\cong\eta_1\oplus\eta_2,\qquad \eta\cong\eta_3\oplus\eta_4,
$$
where $\eta_1$,\dots,$\eta_4$ are the tautological line bundles over
the corresponding factors in $BT^4$.
To calculate the Gysin map in this situation we can use standard
techniques from fixed point theory. The bundle
$CP(\eta_1\oplus\dots\oplus\eta_4)$ has four section
$s_1$,\dots,$s_4$ which are in one-to-one correspondence with the
summands in $\eta_1\oplus\dots\oplus\eta_4$. Let us denote the
first Chern class of $\eta_i$ by $\epsilon_i$. Now for rational
cohomology the complex structure (\ref{tauf}) induces the standard
orientation of the fibre. Thus we obtain
\begin{equation}
p_!(f(\gamma_1)\cdots f(\gamma_4))=\sum^4_{i=1}\frac{s^*_i(f(\gamma_1)\cdots
f(\gamma_4))}{\prod_{j\neq i}(\epsilon_j-\epsilon_i)}.
\label{GYSIN}
\end{equation}
Using the explicit form of the complex structure (\ref{tauf}) in
$\tau_F$ we derive:
$$
s^*_1(\tau_F)=\eta^*_1\otimes (\eta_1\oplus\eta_2)\oplus\eta_1\otimes
(\eta^*_3\oplus\eta^*_4)
\stackrel{\mathbb C}\cong
[1]_{\mathbb C}\oplus\eta^*_1\otimes\eta_2\oplus\eta_1\otimes\eta^*_3
\oplus\eta_1\otimes\eta^*_4;
$$
$$
s^*_2(\tau_F)=\eta^*_2\otimes (\eta_1\oplus\eta_2)\oplus\eta_2\otimes
(\eta^*_3\oplus\eta^*_4)
\stackrel{\mathbb C}\cong
\eta^*_2\otimes\eta_1\oplus [1]_{\mathbb C}\oplus\eta_2\otimes\eta^*_3
\oplus\eta_2\otimes\eta^*_4;
$$
$$
s^*_3(\tau_F)=\eta^*_3\otimes (\eta_1\oplus\eta_2)\oplus\eta_3\otimes
(\eta^*_3\oplus\eta^*_4)
\stackrel{\mathbb C}\cong\eta^*_3\otimes\eta_1\oplus\eta^*_3\otimes\eta_2
\oplus [1]_{\mathbb C}\oplus\eta_3\otimes\eta^*_4;
$$
$$
s^*_4(\tau_F)=\eta^*_4\otimes (\eta_1\oplus\eta_2)\oplus\eta_4\otimes
(\eta^*_3\oplus\eta^*_4)
\stackrel{\mathbb C}\cong\eta^*_4\otimes\eta_1\oplus\eta^*_4\otimes\eta_2
\oplus\eta_4\otimes\eta^*_3\oplus [1]_{\mathbb C}.
$$
Because $f([1]_{\mathbb C})=1$ and $f(c_1(\eta^*_i\otimes\eta_j))=
f(\epsilon_j-\epsilon_i)$ we deduce the following explicit formulae for
the restrictions on the sections $s_j$, $j=1,2,3,4$, of
$CP(\eta_1\oplus\eta_2\oplus\eta_3\oplus\eta_4)$:
$$
s^*_1(f(\gamma_1)\cdots
f(\gamma_4))=f(\epsilon_2-\epsilon_1)f(\epsilon_1-\epsilon_3)
f(\epsilon_1-\epsilon_4);
$$
$$
s^*_2(f(\gamma_1)\cdots
f(\gamma_4))=f(\epsilon_1-\epsilon_2)f(\epsilon_2-\epsilon_3)f(\epsilon_2-\epsilon_4);
$$
$$
s^*_3(f(\gamma_1)\cdots
f(\gamma_4))=f(\epsilon_1-\epsilon_3)f(\epsilon_2-\epsilon_3)f(\epsilon_3-\epsilon_4);
$$
$$
s^*_4(f(\gamma_1)\cdots
f(\gamma_4))=f(\epsilon_1-\epsilon_4)f(\epsilon_2-\epsilon_4)f(\epsilon_4-\epsilon_3).
$$
Finally from~(\ref{GYSIN}), the condition
$p_!(f(\gamma_1)\cdots f(\gamma_4))\in
H^0(BT^4,\mathbb Q)$ is equivalent to the functional equation
(\ref{funl}).
\end{proof}

\section{Solution of the Functional equation}
We shall now obtain the solution to our functional equation.
\begin{Th} Let $g(x)=f(x)/x$. The general analytic solution to
(\ref{funl}) with expansion $g(x)=1/x+\alpha_1+\alpha_2 x+\dots$
is given by the Krichever-H\"ohn elliptic genus
$$g(x)=e\sp{\,\mu\, x}\,\frac{\sigma(\nu-x)}{\sigma(\nu)\,\sigma(x)}.$$
Here $\sigma(x)=\sigma(x|\omega,\omega\sp\prime)$ is the
Weierstrass sigma function which may alternately be expressed in
terms of the Jacobi theta function $\theta_1$ as
$\sigma(x|\omega,\omega\sp\prime)=\frac{2
\omega}{\pi}\exp\left[\frac{\eta
x^2}{2\omega}\right]{\theta_1\left(\frac{\pi x}{2
\omega}\vert\frac{\omega'}{\omega}\right)}/{\theta_1'}$. Moreover,
the only possible value for $\lambda$ in~(\ref{funl}) is zero.
\end{Th}

A particular case of this will be the Ochanine genus, when $g(x)$
is odd (which corresponds to $\nu$ being a half-period), and the
functional equation becomes that studied by Hirzebruch \cite{Hi}.
The connection with (\ref{chiy}) is made using the Jacobi triple
product formula
$$\frac{\theta_1\left(\frac{ix}{2}\,\vert \frac{\omega\sp\prime}{\omega}
\right)}{\theta_1'\left(0\,\vert \frac{\omega\sp\prime}{\omega}
\right)}=i\, \sinh(x/2)\prod_{k=1}\sp{\infty}\frac{(1-{\bar
q}\sp{2k} e\sp{x}) (1-{\bar q}\sp{2k} e\sp{-x})}{(1-{\bar
q}\sp{2k} )^2}, \qquad {\bar q}=\exp\left(i \pi
\frac{\omega\sp\prime}{\omega}\right).$$ Then with $\mu =\eta
\nu/\omega$ we have \begin{align*} g_{\mu =\eta
\nu/\omega}(\frac{i \omega x}{\pi})&=\frac{\theta_1'\left(0\,\vert
\frac{\omega\sp\prime}{\omega}
\right)}{2i\theta_1\left(\frac{i\nu}{2}\,
\vert\frac{\omega\sp\prime}{\omega} \right)}\frac{
\theta_1\left(i[x-\nu]/2\,\vert
\frac{\omega\sp\prime}{\omega}\right)}{\theta_1\left(ix/2\, \vert
\frac{\omega\sp\prime}{\omega}\right)}\\
&=\frac{\theta_1'\left(0\,\vert \frac{\omega\sp\prime}{\omega}
\right)} {2i\theta_1\left(\frac{i\nu}{2}
\,\vert\frac{\omega\sp\prime}{\omega} \right)}
\frac{\sinh([x-\nu]/2)}{\sinh(x/2)}
\prod_{k=1}\sp{\infty}\frac{(1-{\bar q}\sp{2k} e\sp{x-\nu})
(1-{\bar q}\sp{2k} e\sp{-x+\nu})}{(1-{\bar q}\sp{2k} e\sp{x})
(1-{\bar q}\sp{2k} e\sp{-x})},
\end{align*}
which yields (\ref{chiy}) up to a normalisation upon setting
$\tilde q={\bar q}^2$ and $-y=\exp(-\nu)$.

The strategy of our proof will be to first show that (\ref{funl})
is a particular example of the the more general equation
\begin{equation}\label{functional}
\phi_1(x+y)= \frac{\left|\begin{array}{cc}
\phi_2(x)&\phi_2(y) \\
\phi_3(x)&\phi_3(y)\\
\end{array}\right|}{\left|\begin{array}{cc}
\phi_4(x)&\phi_4(y)\\
\phi_5(x)&\phi_5(y) \\
\end{array}\right|}
\end{equation}
studied by Braden and Buchstaber \cite{BB}. This equation includes
many functional equations of cohomological interest. The cited
work in fact provides a constructive method of solution we shall
utilise. The general analytic solution of (\ref{functional}) is,
up to symmetries, given by
$$\phi_1(x)= \frac{\Phi(x;\nu_1)}{\Phi(x;\nu_2)},\quad
 \binom{\phi_2(x)}{\phi_3(x)}=\binom{\Phi(x;\nu_1)}{\Phi\sp{\prime}(x;\nu_1)}\quad
{\rm and}\quad
 \binom{\phi_4(x)}{\phi_5(x)}=\binom{\Phi(x;\nu_2)}{\Phi\sp\prime(x;\nu_2)},
$$
where
\begin{equation}\label{soln}
\Phi(x;\nu)\equiv \frac{\sigma(\nu-x)}{ {\sigma(\nu)\sigma(x)}}\,
  e\sp{\zeta(\nu)x}.
\end{equation}
Here $\zeta(x) =\frac{\sigma(x)\sp\prime}{\sigma(x)}$ is the
Weierstrass zeta function. The parameters appearing in the
solution are determined as follows. Suppose $x_0$ is a generic
point for (\ref{functional}). Then (for $k=1,2$) we have that
\begin{align}
\partial_y \ln
         \left|\begin{array}{cc}\phi_{2k}(x+x_0)&\phi_{2k}(y+x_0)\\
          \phi_{2k+1}(x+x_0)&\phi_{2k+1}(y+x_0)\\
 \end{array}\right|
\Biggl|_{y=0} &= \zeta(\nu_k)-\zeta(x)-\zeta(\nu_k-x)-\lambda_k,
\label{funconst}
\\
&=
 -\frac{1}{ x}-\lambda_k +\sum_{l=0}F_l\, \frac{x\sp{l+1}}{
 (l+1)!}.
 \nonumber
\end{align}
The Laurent expansion determines the parameters $g_1$, $g_2$
(which are the same for both $k=1,2$) characterising the elliptic
functions of (\ref{soln}) by
$$
g_2=\frac{5}{3}\left({F_2+6 F_0\sp2 }\right), \quad\quad g_3= 6
F_0\sp3 -F_1\sp2 +\frac{5}{3}F_0 F_2,
$$
and the parameters $\nu_k$ via $F_0=-\wp(\nu_k)$. Here
$\wp(x)=-\zeta'(x)$ is the Weierstrass elliptic $\wp$-function
with periods $2\omega$, $2\omega'$ that satisfies the differential
equation $\wp'(x)^2=4 \wp(x)^3-g_2 \wp(x)-g_3$.

\begin{proof} Upon setting $g(x)=f(x)/x$ equation (\ref{funl})
may be rewritten as
\begin{align}
\lambda&={g(x_2-x_1)}\,{g(x_1-y_1)}\, {g(x_1-y_2)}+
{g(x_1-x_2)}\,{g(x_2-y_1)}\,{g(x_2-y_2)}\nonumber\\
&\qquad -{g(x_1-y_1)}\,{g(x_2-y_1)}\, {g(y_1-y_2)}-
{g(x_1-y_2)}\,{g(x_2-y_2)}\,{g(y_2-y_1)}, \nonumber \\
&= {g(-a)}\,{g(a+b)}\, {g(a+b+c)}+
{g(a)}\,{g(b)}\,{g(b+c)}\label{funla} \\
&\qquad - {g(a+b)}\,{g(b)}\, {g(c)}-
{g(a+b+c)}\,{g(b+c)}\,{g(-c)},\nonumber
\end{align}
where $x_1-x_2=a$, $x_2-y_1=b$ and $y_1-y_2=c$.

First observe $\lambda=0$. This may be seen by substituting
$g(x)=1/x+\alpha_1+\alpha_2 x+\dots$ into (\ref{funla}) and
determining the constant term. Then from the $b=0$ pole term of
(\ref{funla}) we find that
\begin{equation}\label{funpole}
0=g(a+c)\left[ g(a)g(-a)-g(c)g(-c)\right]+g(a)g'(c)-g'(a)g(c).
\end{equation}
This equation is then of the form (\ref{functional}):
\begin{equation}\label{funcomp}
g(a+c)=\frac{\left|\begin{array}{cc}
  g(a) & g(c) \\
  g'(a) & g'(c) \\
\end{array}\right|}{\left|\begin{array}{cc}
  1 & 1 \\
  g(a)g(-a) & g(c)g(-c) \\
\end{array}\right|}.
\end{equation}

We shall now employ the constructive techniques of \cite{BB}.
Applying (\ref{funconst}) to the denominator of (\ref{funcomp})
yields
\begin{equation}\label{funpdet}
\dfrac{g'(x_0)g(-x_0)-g(x_0)g'(-x_0)}{g(x_0)g(-x_0)-g(x+x_0)g(-x-x_0)}=
\zeta(\nu_2)-\zeta(x)-\zeta(\nu_2-x)-\lambda_2.
\end{equation}
Now the left-hand-side has poles at $x=0$, corresponding to
$\zeta(x)=1/x$ on the right-hand-side, and a further pole at
$x=-2x_0$. From this we deduce that $\nu_2=-2x_0$. Also, from the
pole in $g(x+x_0)$ as $x\rightarrow -x_0$ we obtain
$$ 0=\zeta(-2x_0)-\zeta(-x_0)-\zeta(-x_0)-\lambda_2$$
and so $\lambda_2=2\,\zeta(x_0)-\zeta(2x_0)$. Upon using standard
elliptic function identities we may rewrite (\ref{funpdet}) as
$$
\dfrac{g(x+x_0)g(-x-x_0)-g(x_0)g(-x_0)}{g'(x_0)g(-x_0)-g(x_0)g'(-x_0)}=
\left[\wp(x_0)-\wp(x+x_0)\right]\dfrac{\sigma\sp4(x_0)}{\sigma(2x_0)}=
-\dfrac{\wp(x_0)-\wp(x+x_0)}{\wp'(x_0)}.
$$
Comparison of the $1/(x+x_0)^2$ pole terms in this equation
enables us to deduce that
$$g'(x_0)g(-x_0)-g(x_0)g'(-x_0)=-\wp'(x_0),$$
and so
$$g(x_0)g(-x_0)=\wp(\nu)-\wp(x_0)=\dfrac{\sigma(-\nu+x_0)\,\sigma(\nu+x_0)}{
\sigma\sp2(\nu)\,\sigma\sp2(x_0)}.
$$

Thus we may write
$$g(x)=h(x)\,\frac{\sigma(\nu-x)}{\sigma(\nu)\,\sigma(x)}$$
where
\begin{equation}\label{eqh}  h(x)\,h(-x)=1.
\end{equation}
It will be convenient to express $h$ as
$$h(x)=e\sp{\psi(x)+\zeta(\nu)x},$$
and so $g(x)=e\sp{\psi(x)}\,\Phi(x;\nu)$. We deduce that $\psi$ is
an odd function from (\ref{eqh}).

Thus far we have only derived constraints from the denominator of
(\ref{funcomp}). Substituting our expression for $g$ into
(\ref{funpole}) now yields
\begin{align*}
g(a+c)\left[ g(c)g(-c)-g(a)g(-a)\right]&=
e\sp{\psi(a+c)}\,\Phi(a+c;\nu)\left[\wp(a)-\wp(c)\right]\\
&=\left|\begin{array}{cc}
  g(a) & g(c) \\
  g'(a) & g'(c) \\
\end{array}\right| \\
&=e\sp{\psi(a)+\psi(c)}\,\Phi(a+c;\nu)\left[\wp(a)-\wp(c)\right]\\
&\qquad+
\left[\psi'(c)-\psi'(a)\right]e\sp{\psi(a)+\psi(c)}\,\Phi(a;\nu)\,\Phi(c;\nu).
\end{align*}
Again using standard elliptic function identities this may be
rewritten as
\begin{align*}
e\sp{\psi(a+c)-\psi(a)-\psi(c)}&=1+ \left[\psi'(c)-\psi'(a)\right]
\dfrac{\Phi(a;\nu)\,\Phi(c;\nu)}{\Phi(a+c;\nu)\left[\wp(a)-\wp(c)\right]}\\
&=1+\dfrac{\psi'(c)-\psi'(a)}{\left(\ln\Phi(c;\nu)\right)'-\left(\ln\Phi(a;\nu)\right)'}\\
&=1+\dfrac{\psi'(c)-\psi'(a)}{\zeta(\nu-a)+\zeta(a)-\zeta(\nu-c)-\zeta(c)}.
\end{align*}
Upon taking the  $c\rightarrow\nu$ limit we obtain
$$e\sp{\psi(a+\nu)-\psi(a)-\psi(\nu)}=1,$$
which has solution
$$\psi(a)=\mu a.$$
We have thus obtained the Krichever-H\"ohn elliptic genus
$$g(x)=e\sp{\,\mu\, x}\,\frac{\sigma(\nu-x)}{\sigma(\nu)\,\sigma(x)},$$
so establishing the theorem.
\end{proof}

\noindent{\textbf{Remark.}} We note that (\ref{funpole}) is
equation [4.3.3] in Hirzebruch \cite{H} which was derived under
the hypothesis of invariance under flops; looking at the pole term
$c=0$ in equation (\ref{funpole}) yields
$$
0=g(a)g(a)g(-a)-\left[\alpha_1\sp2-3\alpha_2\right]g(a)-\alpha_2g'(a)+\frac{1}{2}g''(a),
$$
which is [4.3.2] in Hirzebruch \cite{H}.

\noindent{\textbf{Remark.}} \ The vanishing of $\lambda$ shows
that $\phi(\overline {CP}(3))=0$, yielding an alternative proof to
that of \cite{To}.

\noindent
{\bf Acknowledgements.} The authors express deep gratitude to
Prof. V.~M.~Buchtsaber and Prof. B.~Totaro for numerous helpful discussions.
The work of K.~E.~Feldman has been supported by EPSRC grant GR/S92137/01.

\end{document}